\documentstyle[aps,prl,preprint,tighten]{revtex}
\begin{document}
\title{Nonlinear Response of a Kondo system: 
Direct and Alternating Tunneling Currents.}
\author{Y. Goldin\footnote{goldin@bgumail.bgu.ac.il}
  \   and Y. Avishai\footnote{yshai@bgumail.bgu.ac.il}\\
   Physics Department,
   Ben Gurion University of the Negev\\
   Beer Sheva, Israel}
\date{cond-mat/9710085}
\maketitle
\begin{abstract}
Non - equilibrium tunneling current of an 
Anderson impurity system subject to both 
constant and alternating electric fields is studied. 
A time - dependent 
Schrieffer - Wolff transformation 
maps the time - dependent Anderson Hamiltonian onto a 
Kondo one. Perturbation expansion in powers of the Kondo
coupling strength is carried out up to third order, yielding a 
remarkably simple analytical expression for the tunneling
current. It is found that the zero - bias anomaly is suppressed by 
the ac - field. Both dc and the first harmonic are equally enhanced by 
the Kondo effect, while the higher harmonics are relatively small. 
These results are shown to be valid also below the Kondo temperature. 
\end{abstract}
\newcommand{\beq}{\begin{equation}}
\newcommand{\eneq}{\end{equation}}
\newcommand{\bea}{\begin{eqnarray}}
\newcommand{\enea}{\end{eqnarray}}
\newcommand{\si}{\sigma}
\newcommand{\ek}{\epsilon_{k}}
\newcommand{\akd}{a_{k, \sigma}^{\dagger}}
\newcommand{\ak}{a_{k, \sigma}}
\newcommand{\ed}{\epsilon_{d, \si}}
\newcommand{\csd}{c_{d, \sigma}^{\dagger}}
\newcommand{\cs}{c_{d, \sigma}}
\newcommand{\Vkt}{\tilde{V}_{k d}}
\newcommand{\Vk}{V_{k d}(t)}
\newcommand{\ta}{t_{1}}
\newcommand{\Tp}{\hat{T}_{p}}

The Kondo effect in bulk materials has been extensively studied for 
more than three decades (for review, see 
\cite{Hewson:book93,Fulde:reviews}). The impressive
advance in the fabrication of mesoscopic 
tunneling devices opens 
a road to explore the {\em non - equilibrium} Kondo physics. 
Experiments have been reported on crossed - wire tungsten junctions 
\cite{Gregory92}, quenched lithographic point contacts 
\cite{RalphBuhrman},
metal and metallic glass 
break junctions \cite{breakJunctions} and, 
recently, quantum dots \cite{QDsKondoExp}. 
Similar progress has also 
been recorded in numerous theoretical works,
\cite{KondoTheoryNE,SivanWingreen96}. 
Within this realm,  theoretical interest is focused on the
physics of a Kondo system subject to nonlinear 
time - dependent fields 
\cite{HettlerSchoeller95,SchillerHershfield96,Ng96,%
LangrethNordlander94_8}. 
Although an appropriate experimental research 
has not yet been carried out, the required ranges of frequencies 
and temperatures have already been achieved in transport measurements 
so there is no real obstacle to perform the pertinent experiments. 
 
So far, calculations of the current through a Kondo system subject to 
a time - dependent bias were carried out 
using various assumptions and
approximations \cite{HettlerSchoeller95,SchillerHershfield96,Ng96}. 
In the present work we carry out straightforward perturbation
expansion of the current in powers of the coupling constant between the 
impurity (or the quantum dot) and the 
conduction bands. Indeed, perturbation theory proved its usefulness for the
 equilibrium Kondo model right from its onset \cite{EarlyPertub}. 
Recently it was used by Sivan and Wingreen \cite{SivanWingreen96} for
the Anderson model (in the Kondo regime) with a constant voltage bias.

Our strategy is to start from the time - dependent Anderson 
model Hamiltonian and then carry out a time - 
dependent Schrieffer - Wolff transformation to map it onto the
 corresponding Kondo - type Hamiltonian.
The next step is to combine a specific approach suggested by Coleman 
\cite{Coleman84} to 
treat interaction in strongly correlated systems with the 
 Schwinger -- Keldysh non - 
equilibrium Green's functions formalism. 
It makes the problem amenable for perturbation
 expansion which is performed below up to  
third order in the Kondo coupling $ J $ 
(sixth order in the tunneling strength). 
The outcome of this procedure is an  
analytical expression for the tunneling current,
which is used to obtain two novel results pertaining to the 
direct and alternating currents, namely: 1)
The zero - bias anomaly in the 
direct current is smoothed out by an external alternating field. 
2) The zeroth and the first harmonic of the 
time - dependent current are enhanced by the Kondo effect 
while the other harmonics remain relatively small. 

At voltages and frequencies less than the level spacing in the 
tunneling region a convenient way
to describe a tunneling system is the Anderson Hamiltonian: 
\beq                                      \label{AndersonHam}
   H_{A} = \sum_{k \in L,R; \si} \left( \ek + 
                                   \Delta_{L(R)}(t) \right) \akd \ak 
          + \sum_{\si} \ed \csd \cs  
          + \frac{1}{2} U \sum_{\si, \sigma' \ne \si} n_{\si} n_{\si' }
          + \sum_{k \in L,R; \si} \left( \Vkt \akd \cs +
                                           \mbox{h.c.}  \right).  
\eneq
Here $ \akd (\ak) $ creates (annihilates) an electron with momentum
 $ k $ and spin $ \si $ in 
one of the two leads, $ \csd (\cs) $ creates (annihilates) an electron
 with spin $ \si $ in the tunneling region (hereafter we call it ``dot''), 
$ \ek $ and $ \ed $ are one -- particle energies in the leads and in
the dot respectively, $ U $ is the Coulomb
interaction energy in the dot, 
$ n_{\si} \equiv \csd \cs $, while $ \Vkt $ are transfer matrix elements
between the leads and the dot. The external fields are included 
through the potential shifts of the leads $ \Delta_{L(R)}(t) $: 
\beq                                    \label{Delta}
  \Delta_{L(R)}(t)  \equiv  \phi_{L(R)} + 
                               W_{L(R)} \cos \left( \Omega t +
\alpha_{L(R)}
 \right). 
\eneq
The first term describes a dc potential, while the second one is 
due to an ac field, which, for simplicity, is assumed 
to be monochromatic. 
We note that chemical potentials of the leads are shifted by the same 
amount $ \Delta_{L(R)}(t) $ as the one - particle energies, 
so the population of energy levels in the leads does not change. 

Although it is possible to 
carry out a  perturbation expansion within the above
model using the slave 
boson technique \cite{SivanWingreen96} it turns out to be very
cumbersome for 
time - dependent problems. Instead, let us first map the Hamiltonian
 (\ref{AndersonHam}) onto a 
Kondo - like one using a {\em time - dependent version}
of the Schrieffer-Wolff transformation
\cite{SchriefferWolff66}. This procedure 
(whose details will  be explained elsewhere) appears to be 
straightforward albeit not trivial. 
Its application is limited to the
Kondo regime, which is determined by the condition
\beq                                         \label{KondoRegime}
   \ed < 0, \hspace{1.0cm} \ed + U > 0, \hspace{1.0cm} 
     \left| \ed \right|,  \ed + U  \gtrsim \Gamma_{\si}, 
\eneq
where 
$ \Gamma_{\si} =  2 \pi  \sum_{k \in L,R} \left| \Vkt \right|^{2}
\delta (\ed - \ek) $ are the widths of the energy levels in the dot. 
Furthermore, it is assumed that the external fields are not strong 
enough to draw the system out of this regime, so that: 
\beq                                                 \label{ExtFcond}
    \left| \phi_{L(R)} \right| \lesssim \left| \ed \right|, \left| \ed +
 U \right|  
       \hspace{1.0cm} 
    \Omega, W_{L},W_{R} \lesssim \left| \ed \right|, \left| \ed + U \right|.  
\eneq
It should be stressed however that these conditions 
do not imply a linear response.
The later is defined by the conditions 
$ \left| \phi_{L(R)} \right|, W_{L(R)} \ll T $ while 
$ T \ll \left| \ed \right|, \left| \ed + U \right|  $. 
For the sole purpose of having a more compact 
form of the Hamiltonian the 
 infinite $ U $ limit is taken. 
The resulting Hamiltonian then reads,
\beq                                     \label{KondoHamFnl}
   H_{K} =  \sum_{k \in L,R; \si} \ek  \akd \ak  +   
      1/2 \! \! \sum_{k, k' \in L,R; \si} \! \! J_{k'k}(t) 
        \left[ 	a_{k', -\si}^{\dagger} \ak \csd c_{d, -\si} + 
	a_{k', \si}^{\dagger} \ak \csd c_{d, \si}   \right],   
\eneq
where 
\bea                                     \label{Jkpk}
   J_{k'k}(t) &  = & V_{k'd} V_{kd}^{*} 
          \exp \left[ \left( \phi_{(k')} - \phi_{(k)} \right) t \right] 
       \sum_{s', s = -\infty}^{+\infty} 
         J_{s'}(\frac{W_{(k')}}{\Omega}) J_{s}(\frac{W_{(k)}}{\Omega}) 
   \exp \left[ (s' -s) \Omega t + i (s' \alpha_{(k')} -s \alpha_{(k)}) \right]
                                      \cdot     \nonumber  \\
 & & \cdot \left( \frac{1}{\epsilon_{k'} + \phi_{(k')} + s' \Omega - \ed}+ 
                       \frac{1}{\ek + \phi_{(k)} + s \Omega - \ed}   
                                          \right),  
\enea
$ V_{kd} \equiv \tilde{V}_{kd} 
            \exp \left[ - i \left(W_{(k)} / \Omega \right) \sin \alpha_{(k)}
                        \right]  $, 
the symbol $ (k) $ means ``L'' or ``R'' depending whether $ k $ belongs 
to the left or to the right lead, 
while $ J_{s}(\frac{W}{\Omega}) $ are Bessel's functions. 

Before proceeding with the evaluation of the current, two 
comments are in order:
i) Any procedure toward solution of the ensuing transport equations 
should take into account the fact that, out of the full Hilbert space, the 
system is projected onto a subspace $ F_{1} $
for which the dot is occupied by one (and only one) electron. 
ii) At this stage one might be tempted to 
express the electron creation -
 annihilation operators in the dot through  
spin operators \cite{SchriefferWolff66}, 
thus arriving at the familiar form 
\cite{Hewson:book93,SchriefferWolff66} of the Kondo Hamiltonian. 
But then one would  realize that the spin operators do not obey the 
usual commutation rules. In order to overcome
this obstacle, fictitious (auxiliary) fermions 
\cite{Abrikosov65} might be introduced. 
But this leads one back to equation
 (\ref{KondoHamFnl}). 
In other words, auxiliary fermions 
which sometimes regarded as artificial 
particles introduced to represent spins are real
 electrons 
in the dot (impurity atom) subject to the constraint specified in i).

Calculation of tunneling current starting from the Kondo Hamiltonian 
(\ref{KondoHamFnl})
is possible for arbitrary field strengths and frequency provided the 
inequalities (\ref{ExtFcond}) are satisfied. Yet, inspecting 
a typical experimental setup \cite{QDsKondoExp} 
one may consider somewhat weaker external fields and lower 
frequencies, so that 
$ \left| \phi_{L(R)} \right|, \Omega, W_{L},W_{R} \ll \left| \ed \right| $. 
Expression (\ref{Jkpk}) for $ J_{k'k} $ then significantly simplifies. 
Matrix elements $ J_{k'k} $ become time - independent if $ k $ and $ k' $ 
belong to the same lead. Moreover, $ J_{k'k} $ do not depend 
on potential shifts of every lead separately but only on their difference, 
$   \Delta_{LR} \equiv \Delta_{L} - \Delta_{R} \equiv 
	\phi^{dc} + W \cos \left( \Omega t + \alpha \right) $. 
For simplicity, we further assume that $ J_{k'k} $ depend only on 
the leads to which $ k' $ and $ k $ belong, 
independently of the values of $ k' $ and $ k $. 

The current $ I(t) $ is defined as the rate of 
change in the number of electrons
 in a lead. Within the interaction picture, the
commutator of the number operator with the Hamiltonian
 (\ref{KondoHamFnl}) yields an expression for the current,
\bea					\label{curr-CTP}
   I(t) & = & \frac{e}{\hbar} \sum_{k' \in L, k \in R; \si} \! 
        \mbox{Im} \left\{ J_{k'k}(t) 
	\mbox{Tr}_{F_{1}} \left[  \rho_{0} 
	\Tp ( a_{k', -\si}^{\dagger}(t) \ak(t) \csd(t)
 c_{d, -\si}(t) S_{p} ) \right] + 
				\right.	\nonumber \\ 
         & & \hspace{2cm} \left. 
       \mbox{} +  J_{k'k}(t)  \mbox{Tr}_{F_{1}} \left[  \rho_{0} 
	\Tp ( a_{k', \si}^{\dagger}(t) \ak(t) \csd(t) c_{d, \si}(t)
S_{p} )
 \right]
					\right\}, 
\enea
where $ \rho_{0} $ is the initial (equilibrium) density matrix,
while  $ S_{p} $ and $ \Tp $ are, respectively, 
the S - matrix and the time - ordering operator on a closed
 time - path.

We now carry out a perturbation expansion of the above
 expression in powers of the 
coupling strength $ J_{k'k} $ using the Schwinger -- Keldysh non -
 equilibrium Green's 
functions technique. In order to get rid of the constraint to
 the subspace $ F_{1} $ we 
combine it with the method suggested by Coleman \cite{Coleman84}
 for the analogous 
problem in the Anderson model. The perturbation diagrams are drawn
 in Fig.~\ref{diagrams}. Every  
line corresponds to a 
$ 2 \times 2 $ matrix of Green's functions (details are to
 be published elsewhere). 
It is important to note that the Hamiltonian (\ref{KondoHamFnl})
 (and, consequently, the 
expression (\ref{curr-CTP}) for the current) contains not
 only the s -d coupling but also the 
usual potential scattering terms (see Ref. \cite{Hewson:book93})
 for explanation). As a result, all 
possible combinations of $ \si $, $ \si_{1} $ and $ \si_{2} $ 
(i.e. $ \si_{1} = \pm \si $, $ \si_{2} = \pm \si $) are
 allowed for diagram A but only one 
of them (namely, $ \si_{2} = \si_{1} = \si $) is allowed
 for diagram B. The spin - flip 
diagrams of the type B cancel. 

The resulting current is given by the following formulae:
\bea              			\label{curr_fnl} 
  I(t) & = & I^{(2)}(t) + I^{(3)}(t)                      \\ 
  I^{(2)}(t) & = & C_{2} \left[ \phi^{dc} + W \cos 
(\Omega t + \alpha) \right] 
					\nonumber       \\
  I^{(3)}(t) & = & \frac{1}{2} I_{0} +  
		     \sum_{n=1}^{+\infty} \left| I_{n} \right| 
			 \cos (n \Omega t + n \alpha 
+ \arg I_{n} ),  \nonumber    \\
  I_{n} & = & C_{3} \sum_{s = -\infty}^{+\infty}  
	J_{s}(\frac{W}{\Omega}) J_{s+n}(\frac{W}{\Omega}) 
	\left[ F ( \phi^{dc} + s \Omega, T, D ) + 
		(-1)^{n} F^{*}( \phi^{dc} - s \Omega, T, D ) \right] , 
								\nonumber  \\
  F( \phi, T, D ) &  =  & - \mbox{Re} \int \!\! 
\int_{-D}^{+\infty} d\omega \, d\epsilon \,
		\frac{ \left[ f_{L}(\omega) - f_{R}(\omega) \right] 
			\left[ f_{L}(\epsilon) + f_{R}(\epsilon) \right] }
		        {\omega - \epsilon + i \gamma } 
	-  i \pi \frac{\phi}{2} \coth \frac{\beta \phi}{2} , 
							          \nonumber  
\enea
where $ I^{(2)} $ and $ I^{(3)} $ 
express contributions of second (diagram C) and  
third (diagrams A and B) orders in $ J_{k'k} \rho $, while
$ C_{2} \equiv \frac{e}{\hbar} \pi 
	\left| \tilde{J}_{LR} \right|^{2} \rho_{L} \rho_{R} $ and 
$ C_{3} \equiv \frac{e}{\hbar} 5 \pi 
	\left| \tilde{J}_{LR} \right|^{2} \rho_{L} \rho_{R} 
	\left( \tilde{J}_{LL} \rho_{L} + \tilde{J}_{RR} \rho_{R}
\right) $. The quantities
$ \rho_{L(R)} $ are densities of states in the leads, 
whereas $ J_{L(R)R(L)} = V_{L(R)} V^{*}_{R(L)} / | \ed | $. 
The cutoff $ D $ is equal to the energy difference between the chemical 
potential and the bottom of the conduction band, while 
$ f_{L}(\epsilon) \equiv 
1 / \left( \exp [ ( \epsilon - \phi )/ k T] + 1 \right) $ and 
$ f_{R}(\epsilon) \equiv 1 / \left( \exp [ \epsilon / k T] +1 \right) $ 
are Fermi functions in the leads (the left lead being shifted by $ \phi $). 
The integral in the expression for $ F $ can be written in 
the same form as in the dc result of Sivan 
and Wingreen \cite{SivanWingreen96}. The 
alternating field causes splitting of the 
energy levels in one of the leads \cite{comment1}. 
Therefore the time - dependent current is 
a result of interference between ``dc - like'' 
contributions, each one with an effective bias $ \phi^{dc} \pm s \Omega $. 

An approximate evaluation of the double 
integral in the above formulae is possible both for the linear 
($ \phi \ll T $) and for the non - 
linear ($ \phi \gg T $) regimes. The result is 
\beq				
				\label{Fapprox}
  F( \phi, T, D )  =  \left[ \begin{array}{c}            
			  \phi  \ln \frac{D}{kT}, 
\mbox{ if $ \phi \ll T $}  \\                   
                      	\phi  \ln \frac{D}{|\phi|}, 
\mbox{ if $ \phi \gg T $}  
                         \end{array}      \right.  .
\eneq
The terms that are not included in this 
expression are of order  $ \phi $. It should be stressed
that in the nonlinear case on which we focus our attention here the 
function $ F $ does not diverge as $ T \rightarrow 0 $. Hence, our 
results for the nonlinear response 
must be valid even below the Kondo temperature. 
This results from the fact that here, the non - 
linear bias plays the role of temperature as the 
largest low - energy scale. 

Substitution of expression (\ref{Fapprox}) 
into equations (\ref{curr_fnl}) yields simple formulae 
for the current. First of all, in the absence of an ac - field we obtain, 
\beq                                      \label{IdcApprox}
  I  \approx  \phi^{dc} \cdot \left[ \begin{array}{c}            
       C_{2} + C_{3} \ln \frac{D}{kT}, \mbox{ if $ \phi^{dc} \ll T $}, \\                   
       C_{2} + C_{3} \ln \frac{D}{| \phi^{dc}|}, 
                                     \mbox{ if $ \phi^{dc} \gg T $}.  
                         \end{array}      \right. 
\eneq
We note that, to the best of our knowledge, such a simple expression for 
a non - equilibrium tunneling current through a Kondo system has not been 
found before. In the presence of a strong ($ W \gg \Omega $) ac - field 
we get, 
\bea                                \label{curr_simple}
  I(t) & \approx & I^{dc} + I^{ac}(t) \\
  I^{dc} & = & \phi^{dc} 
	\left[ C_{2} + C_{3}  
	\ln \frac{D}{ A_{dc} \left( \phi^{dc},W, \Omega, T \right) } \right], 
					\nonumber \\
  I^{ac}(t) & = & W \cos (\Omega t + \alpha) 
	\left[ C_{2} + C_{3}  
	\ln \frac{D}{ A_{ac} \left( \phi^{dc},W, \Omega, T \right) } \right], 
					\nonumber 
\enea
where $ A_{dc}, A_{ac} $ are of the order of 
$ \max(\phi^{dc},W) $, if $ \phi^{dc} \gg T $ or $ W \gg T $ 
(nonlinear response), and 
$ A_{dc} = A_{ac} = T $, if $ \phi^{dc} \ll T $ and $ W \ll T $ 
(linear response). 

In most of the relevant experiments the attention was focused on the dc. 
The best pronounced feature of the Kondo effect in the dc seems to be the 
zero - bias anomaly, i.e. appearance of a narrow peak in the differential 
conductance around zero bias. In Fig. \ref{ZBA_A} and \ref{ZBA_B} 
we show how it is altered by application of the external time - dependent 
field. According to formula (\ref{curr_fnl}), dc in an external ac - field is 
given by a sum of direct currents without ac - field but with effective 
biases $ \phi^{dc} + s \Omega $ (these currents are weighted by 
$ J_{s}^{2} $). Consequently, the zero - bias anomaly is suppressed. A 
picture of gradual flattening similar to Fig. \ref{ZBA_A} holds also at 
lower frequencies down to the adiabatic limit. At higher frequencies, 
however, side peaks at multiples of $ \Omega $ can be resolved (Fig. 
\ref{ZBA_B}). 

Although measurement of an alternating 
current with frequencies 
and amplitudes in the relevant range is not an easy task, 
it might reveal new interesting features of 
the Kondo effect. Expression (\ref{curr_simple}) shows that only the dc 
and the first harmonic are enhanced by the Kondo effect. 
For the other harmonics, the interference of the 
contributions to the current in equations (\ref{curr_fnl}) 
with different effective biases is destructive. 
We emphasize that the nonlinear response 
is different from the linear one. Namely, 
(1) the second and the higher harmonics exist 
although they are not amplified by the large factor $ \ln \frac{D}{A} $, 
(2) the factor $ \ln \frac{D}{T} $ is replaced by $ \ln \frac{D}{A} $. 
Nevertheless, they {\em look} very similar 
in the sense that the dc and the first harmonic 
are much larger than the others. 
This is a particular feature of the interacting (Kondo) system. In 
the non - interacting one - level system the amplitude of the higher 
harmonics $ I_{n}$, $ n \geq 2 $ is, generally,  
of the order of the amplitudes of the dc and the first one 
(gradually decreasing with their number as $ \Gamma / ( n \Omega ) $). 
In some sense the Kondo system, in contrast
with the non - interacting one, behaves like an 
ordinary resistor (although the current is enhanced by the Kondo effect). 
Unlike the situation encountered in non - interacting systems,
the magnitudes of the dc 
and the ac components are mostly governed by two independent 
parameters $ \phi^{dc} $ and $ W $. 
Experimentally, Kondo contribution to the tunneling current is usually 
revealed through a special dependence on the parameters (such as 
$ \ln T $ increase of the conductance or zero - bias anomaly). Our 
formulae (\ref{curr_simple}) imply that this kind of effects can be found 
in the first harmonic as well as in the dc but not in higher harmonics. 

In conclusion, the pertinent physics is rather rich 
since it combines strong correlations with nonlinearity 
and time dependence. 
We obtain (within perturbation theory) 
rigorous analytical formulae for the nonlinear time - dependent 
tunneling current in the Kondo regime. They apply for nonlinear 
response both below and above the Kondo 
temperature (for linear response 
they are valid only above it). In the special case of constant 
external field the $I - \phi $ relation (\ref{IdcApprox}) 
encodes the main physics of the familiar zero - bias anomaly.

The first novel result of the present research 
can be easily tested experimentally, 
since it concerns the {\em direct} tunneling current. 
We find that its zero - bias anomaly 
is suppressed by an ac - field. 
Second, it is shown that 
both the zeroth and the first harmonics 
of the alternating current are strongly enhanced by the 
Kondo effect, while the other harmonics remain relatively small. 
This result is slightly more difficult to be tested experimentally. 
Yet, it is apparently worth the effort since it
is remarkably different 
from what is expected for a non - interacting one - level system where 
all the harmonics emerge together. 

As far as relation to previous relevant works is concerned, we, first, 
notice 
that our analytical results for the dc are consistent with the numerical 
calculations of Ref. \cite{HettlerSchoeller95}. However, being able to 
consider 
stronger ac - fields (larger ratio $ W / \Omega $), we find also an overall 
suppression of the zero - bias anomaly, besides the appearance of side 
peaks. As for the spectrum of the tunneling current, 
we are unable to validate 
the assumption suggested in Ref.
\cite{HettlerSchoeller95} that 
{\em all} the harmonics beside the dc one 
can be neglected. On the other hand, we verify that the second and 
higher harmonics are generated but they are indeed much smaller 
than the dc and the first one \cite{comment2}. 
Within a specific model, some authors \cite{SchillerHershfield96} 
obtained current spectrum similar to that of a non - 
interacting system. We attribute the difference between 
this result and ours to the
very peculiar choice of parameters used therein. 

{\em Acknowledgments.} This research was supported in part by 
a grant from the Israel Academy of 
Science and Humanities under {\em Centers of Excellence Program}.  
We would like to thank Y. Meir, A. Golub, N.S. Wingreen, P. Coleman, 
K.A. Kikoin, G. Sch\"{o}n, D. Goldhaber - Gordon and E. Kogan for 
many helpful discussions and comments.

\newpage
\begin{figure}[t]
\caption{Diagrams for the perturbation expansion of the current
(\ref{curr-CTP}). Solid lines stand for lead electrons, dashed lines ---
for dot electrons (auxiliary fermions).\label{diagrams}}
\caption{Suppression of the zero - bias anomaly in the direct current by 
an external alternating field at relatively low frequencies. 
Differential conductance 
$ d I^{dc} / d \phi^{dc} $ is given in units of $ C_{3} T $, while 
$ D = 50 $, $ \hbar \Omega = 5 $, $ W $ and $ \phi^{dc} $ 
are in units of $ T $. 
\label{ZBA_A}}
\caption{Suppression of the zero - bias anomaly in the direct current by 
an external alternating field at higher frequencies. Side peaks appear at 
multiples of $ \Omega $. Differential conductance 
$ d I^{dc} / d \phi^{dc} $ is given in units of $ C_{3} T $, while 
$ D = 200 $, $ \hbar \Omega = 15 $, $ W $ and $ \phi^{dc} $ 
are in units of $ T $. 
\label{ZBA_B}}
\end{figure} 


\begin{thebibliography}{99}
\bibitem{comment2} It is true also below the Kondo temperature. 
We explained it in the text for the nonlinear case. In the linear 
response the function $ F( \phi, T, D ) $ is linear in $ \phi $ by definition. 
Then it can be exactly proved that only 
the dc and the first harmonic are generated.
\end{thebibliography}
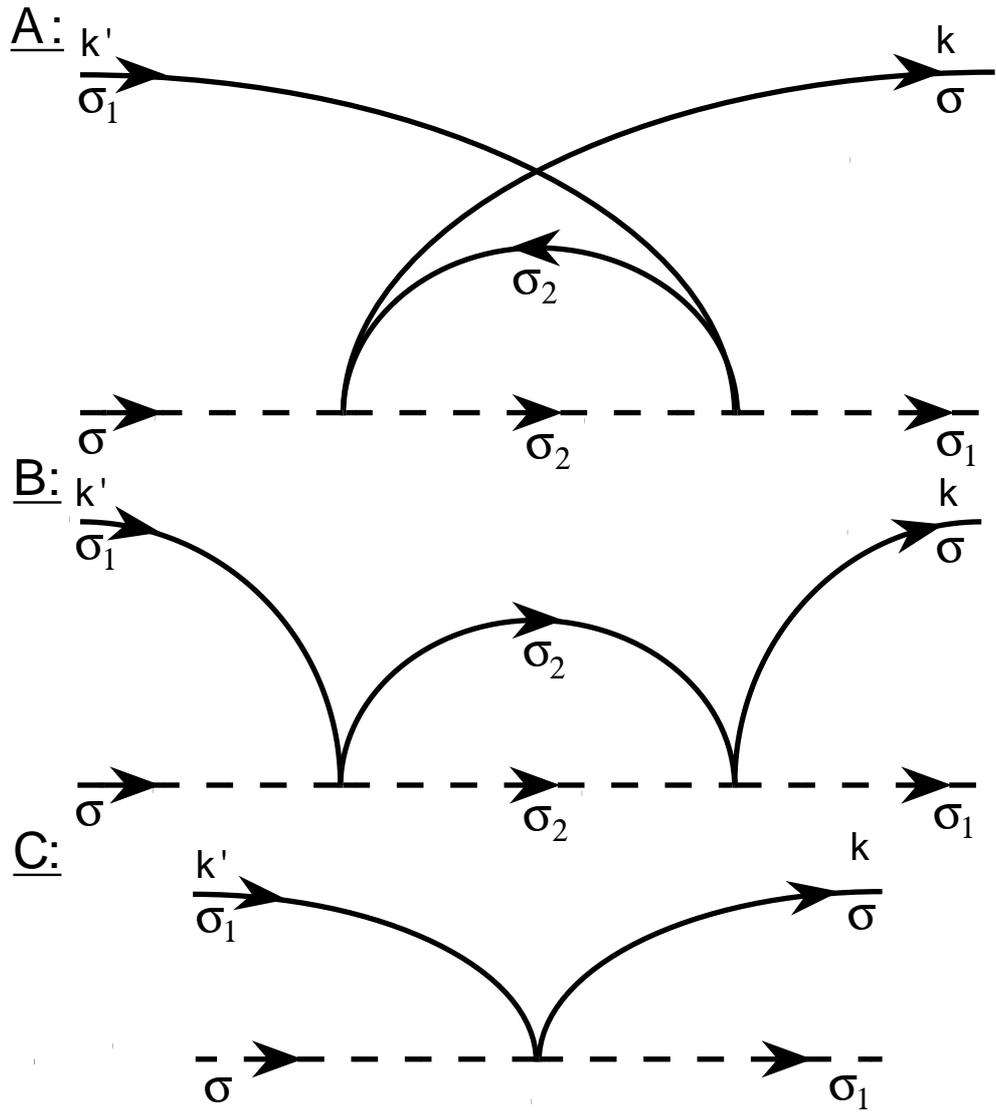
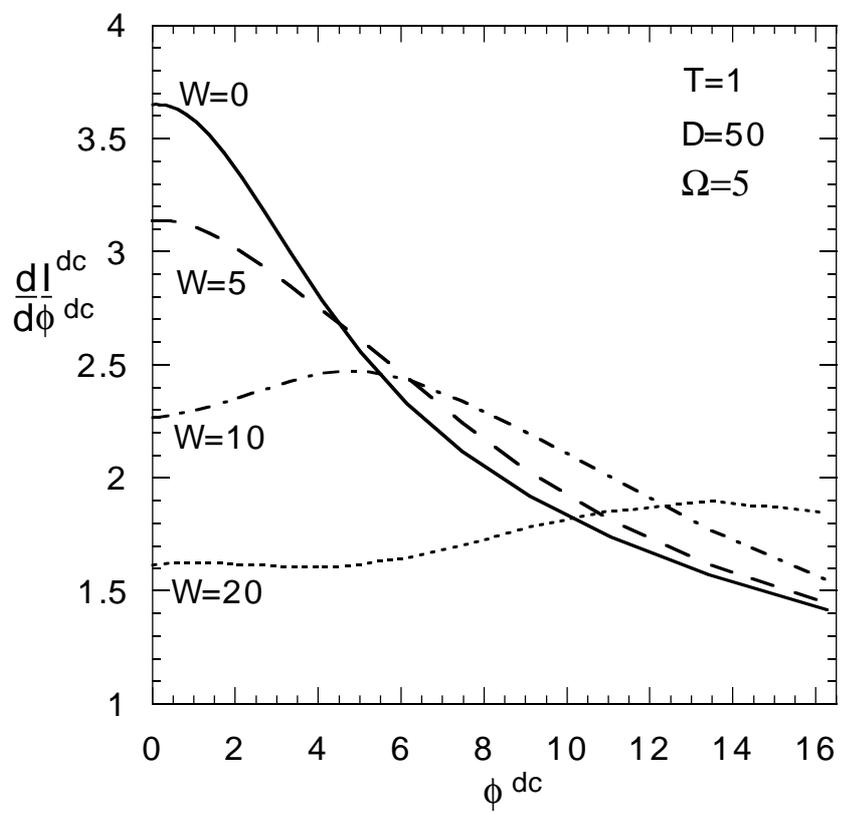
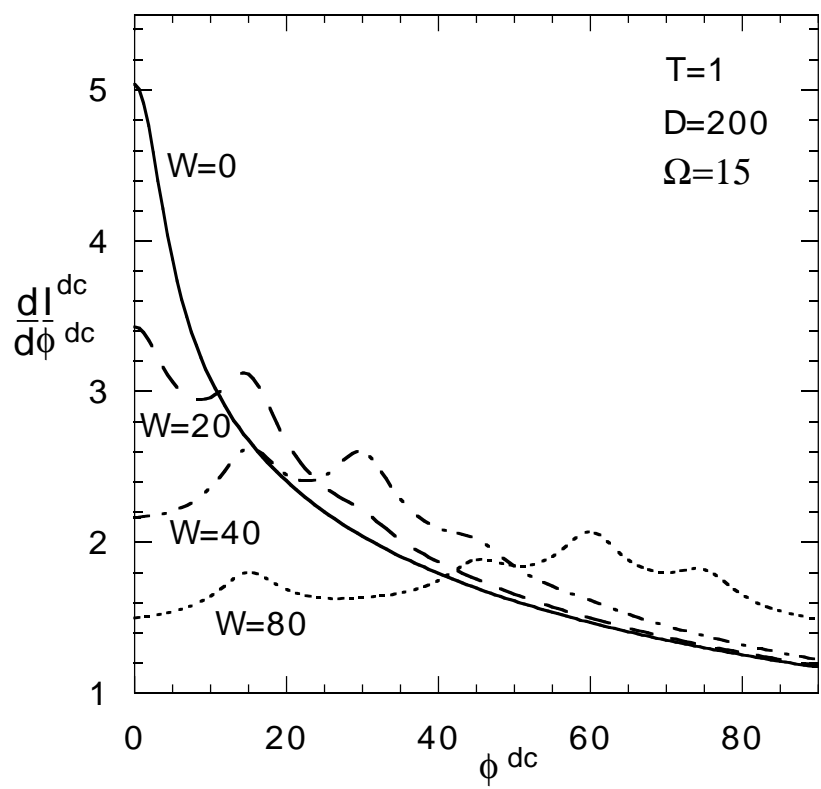
\end{document}